\documentclass[conference]{IEEEtran}
\IEEEoverridecommandlockouts
\usepackage{cite}
\usepackage{amsmath,amssymb,amsfonts}
\usepackage{textcomp}
\usepackage{xcolor}

\usepackage{algorithmic}
\usepackage{algorithm}
\usepackage{booktabs} % 必须引入这个包来画三线表
\usepackage{multirow}
\usepackage{setspace}
\usepackage{amsmath} %数学公式
\usepackage{amssymb}    % AMS数学符号 (例如 \mathbb)
\usepackage[T1]{fontenc} % 字体编码
\usepackage{graphicx}
\usepackage{bm}
\def\BibTeX{{\rm B\kern-.05em{\sc i\kern-.025em b}\kern-.08em
    T\kern-.1667em\lower.7ex\hbox{E}\kern-.125emX}}

\begin{document}

\title{Vector Quantized-Aided XL-MIMO CSI Feedback with Channel Adaptive Transmission}
\author{
    \IEEEauthorblockN{Yuhang Ma, Nan Ma, Jianqiao Chen and Wenkai Liu}
    % \IEEEauthorblockA{...} % 您可以留空，或者在这里写主要的单位
    
    % --- 使用 \thanks 命令添加左下角信息 ---
    % 每一条 \thanks 都会换一行
    
    \thanks{Yuhang Ma, Nan Ma and Wenkai Liu are with the State Key Laboratory of Networking and Switching Technology, Beijing University of Posts and Telecommunications, Beijing 100876, China. Jianqiao Chen is with the ZGC Institute of Ubiquitous-X Innovation and Applications, Beijing 100876, China. Email: \{mayuhang, manan, liuwenkai\}@bupt.edu.cn, jqchen1988@163.com}
    
    %\thanks{通讯作者: 作者一 (author1@example.com).} % 通讯作者信息
}

\maketitle

\begin{abstract}
Efficient channel state information (CSI) feedback is critical for 6G extremely large-scale multiple-input multiple-output (XL-MIMO) systems to mitigate channel interference. However, the massive antenna scale imposes a severe burden on feedback overhead. Meanwhile, existing quantized feedback methods face dual challenges of limited quantization precision and insufficient channel robustness when compressing high-dimensional channel features into discrete symbols. To reduce these gaps, guided by the deep joint source-channel coding (DJSCC) framework, we propose a vector quantized (VQ)-aided scheme for CSI feedback in XL-MIMO systems considering the near-field effect, named VQ-DJSCC-F. Firstly, taking advantage of the sparsity of near-field channels in the polar-delay domain, we extract energy-concentrated features to reduce dimensionality. Then, we simultaneously design the Transformer and CNN (convolutional neural network) architectures as the backbones to hierarchically extract CSI features, followed by VQ modules projecting features into a discrete latent space. The entropy loss regularization in synergy with an exponential moving average (EMA) update strategy is introduced to maximize quantization precision. Furthermore, we develop an attention mechanism-driven channel adaptation module to mitigate the impact of wireless channel fading on the transmission of index sequences. Simulation results demonstrate that the proposed scheme achieves superior CSI reconstruction accuracy with lower feedback overheads under varying channel conditions.
\end{abstract}


\IEEEpeerreviewmaketitle

\section{Introduction}
Recently, extremely large-scale multiple-input multiple-output (XL-MIMO) has emerged as a pivotal enabler for 6G networks, significantly enhancing spatial multiplexing gain \cite{1}. In FDD systems, accurate downlink channel state information (CSI) feedback is indispensable for pre-coding, yet the massive antenna scale imposes severe feedback overhead. Furthermore, as apertures extend to hundreds of wavelengths, the communication model shifts to the radiating near-field \cite{2}. This introduces spherical wave propagation, causing angular energy diffusion and invalidating the sparsity assumption of traditional methods \cite{3}. Consequently, efficient compression of high-dimensional near-field CSI to reduce overhead has become a critical challenge.

In recent years, deep learning (DL)-based methods have gradually superseded traditional compressed sensing (CS) techniques. Early works like CsiNet \cite{4} and other compressive sensing-like schemes \cite{5}, effectively compressed far-field CSI by exploiting angular sparsity. However, in XL-MIMO near-field channels, the inherent spherical wave propagation causes severe energy diffusion, rendering such angular-domain strategies ineffective. To address this, polar domain representations have been introduced to recover sparsity in the near-field \cite{6}. Nevertheless, most existing schemes assume ideal feedback links, ignoring the impact of fluctuating signal-to-noise ratios (SNR). While joint source-channel coding (JSCC) principles \cite{7} can combat channel noise, standard JSCC architectures typically output continuous floating-point features. This analog feedback mechanism is fundamentally incompatible with modern digital communication protocols that rely on discrete bit streams. To bridge this gap, vector quantization (VQ)-based frameworks, following the seminal VQ-VAE \cite{8} and its hierarchical variants \cite{9}, have been adopted in semantic communications \cite{10} and massive MIMO feedback \cite{11}, \cite{12} to ensure intrinsic digital compatibility.

Despite the immense potential of VQ-VAE, several critical aspects require further investigation. First, insufficient robustness against noisy feedback channels remains a major challenge. Since standard VQ mechanisms do not account for channel noise, the corruption of transmitted discrete indices in low SNR regimes can lead to a catastrophic degradation in reconstruction accuracy \cite{13}. Therefore, it is essential to design mechanisms that protect indices and adapt to time-varying channel conditions. Second, the training of VQ architectures is intrinsically hindered by the non-differentiable nature of the quantization layer. Although the straight-through estimator (STE) \cite{14} offers a widely used gradient approximation solution, this approach often suffers from the ``codebook collapse'' problem \cite{15}. In this phenomenon, a significant portion of the codebook vectors remain unutilized, thereby severely constraining the quantization precision. Consequently, developing a VQ-based framework tailored for near-field CSI feedback tasks that simultaneously adapts to channel variations and mitigates codebook collapse is imperative.

To reduce these gaps, we propose a VQ-aided scheme for CSI feedback in XL-MIMO systems within DJSCC framework, termed VQ-DJSCC-F. Our work aims to achieve efficient discretization of near-field features while ensuring reliable index transmission over varying channel conditions. The main contributions are summarized as follows:

\noindent\llap{\textbullet}\hspace{1.1em}\hangindent=1.5em \hangafter=1
\textit{Vector Quantization-aided CSI Feature Extraction and Quantization:} We propose hierarchical Transformer and CNN backbones integrated with VQ modules to project features into a discrete codebook. By leveraging near-field sparsity in the polar-delay domain, we significantly reduce the dimensionality of the original CSI. This architecture efficiently bridges the gap between multi-scale spatial feature extraction and digital transmission, facilitating efficient near-field CSI feedback.

\noindent\llap{\textbullet}\hspace{1.1em}\hangindent=1.5em \hangafter=1
\textit{Codebook Optimization Scheme and Channel Adaptation:} To address the ``codebook collapse'' issue and enhance quantization precision, we introduce entropy loss regularization alongside an exponential moving average (EMA) update strategy to enforce uniform codebook distribution. Furthermore, to mitigate the impact of varying channel conditions on index feedback, we develop an attention-driven adaptation module integrated into the DJSCC framework for robust end-to-end transmission.

The rest of this paper is organized as follows. Section II provides an overview of the system model and problem formulation. Section III elaborates the proposed VQ-enabled CSI feedback scheme for XL-MIMO systems. Section IV presents the simulation results that validate the performance of our proposed method. Section V concludes the paper.

\section{\scalebox{0.9}{SYSTEM MODEL AND PROBLEM FORMULATION}}
As shown in Fig. 1, the proposed VQ-DJSCC-F framework establishes an end-to-end digital feedback link. The architecture consists of three cascaded stages: near-field CSI pre-processing, VQ-based feature compression with index transmission, and CSI reconstruction.

\begin{figure}[htbp]
    \centering
    \includegraphics[width=\linewidth, keepaspectratio]{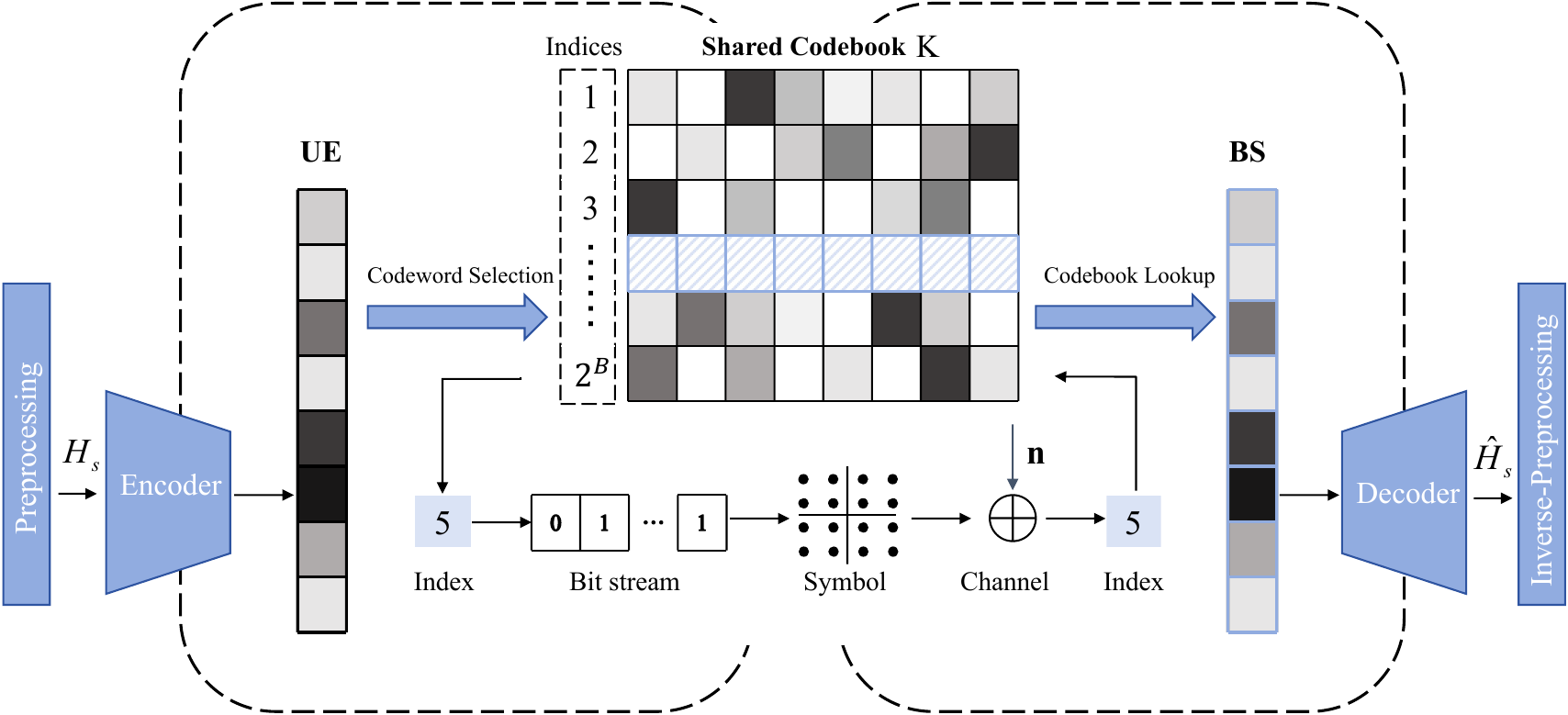}
    \caption{The end-to-end framework of the proposed VQ-DJSCC-F.}
    \label{fig:system}
\end{figure}

The process commences with pre-processing to achieve the sparsity of near-field CSI in the polar-delay domain. Let $\mathbf{H} \in \mathbb{C}^{N_{tx} \times N_{s}}$ denote the original downlink spatial-frequency domain CSI, where $N_{tx}$ and $N_{s}$ represent the number of transmit antennas and subcarriers, respectively. To harness the joint sparsity, we perform domain transformation followed by energy truncation. By retaining only the dominant multipath components, this step achieves preliminary compression of the high-dimensional channel. Consequently, the sparse feature matrix $\mathbf{H}_s \in \mathbb{C}^{N_p \times N_d}$ is obtained.

The compressed sparse matrix $\mathbf{H}_s$ is fed into the VQ-based feedback framework. To enable the system with adaptability to dynamic channel conditions, the encoder $\mathcal{E}_{\theta}(\cdot)$ takes both the sparse CSI and the estimated SNR $\gamma$ as inputs. The latent feature tensor $\mathbf{Z}$ can be expressed as
\begin{equation}
    \mathbf{Z} = \mathcal{E}_{\theta}(\mathbf{H}_s, \gamma),
\end{equation}
where $\theta$ denotes the learnable parameter set of the encoder.

Subsequently, the VQ module discretizes the latent feature tensor $\mathbf{Z}$. We consider $\mathbf{Z}$ as a sequence of $M$ feature vectors, denoted as $\mathbf{Z} = [\mathbf{z}_1, \dots, \mathbf{z}_M]^T \in \mathbb{R}^{M \times d}$, where $d$ represents the feature dimension. The VQ module employs a shared learnable codebook $\mathcal{C} = \{\mathbf{c}_k\}_{k=1}^K \in \mathbb{R}^{K \times d}$, where $K$ denotes the codebook size. The quantization process maps each input vector $\mathbf{z}_{m}$ to the index of its nearest neighbor in the codebook, formulated as
\begin{equation}
    q_m = \underset{k \in \{1, \dots, K\}}{\arg\min} \|\mathbf{z}_{m} - \mathbf{c}_k\|_2^2,
\end{equation}
where $q_m$ represents the optimal codeword index for the $m$-th feature vector. Consequently, the feature matrix $\mathbf{Z}$ is converted into a discrete index sequence $\bm{q} = [q_1, \dots, q_M]^T \in \{1, \dots, K\}^M$.

After modulation, the index sequence $\bm{q}$ is processed as a complex-valued signal $\mathbf{x}$ for transmission. Considering an additive white gaussian noise (AWGN) channel, the received signal $\mathbf{y}$ at the BS can be expressed as
\begin{equation}
    \mathbf{y} = \mathbf{x} + \mathbf{n},
\end{equation}
where $\mathbf{n}$ is the additive noise vector with independent and identically distributed entries following $\mathcal{CN}(0, \sigma^2 \mathbf{I})$.

At the BS, a series of signal processing steps are performed to recover the transmitted information. Based on $\hat{\bm{q}}$, the corresponding quantized feature vectors are retrieved from the codebook $\mathcal{C}$ to reconstruct the latent feature tensor $\hat{\mathbf{Z}}_q$. Finally, by leveraging the feedback SNR $\gamma$ to facilitate reconstruction, the recovered CSI matrix output by the decoder $\mathcal{D}_{\phi}(\cdot)$ can be expressed as
\begin{equation}
    \hat{\mathbf{H}}_s = \mathcal{D}_{\phi}(\hat{\mathbf{Z}}_q, \gamma),
\end{equation}
where $\phi$ denotes the learnable parameters of the decoder.

The core objective is to design a jointly optimized end-to-end system that minimizes reconstruction distortion. Given the stochastic nature of wireless channels, the optimization problem can be formulated as
\begin{equation}
    (\theta^*, \phi^*, \mathcal{C}^*) = \underset{\theta, \phi, \mathcal{C}}{\arg\min} \; \mathbb{E}_{\gamma \sim p(\gamma)} \mathbb{E}_{\mathbf{H}_s \sim p(\mathbf{H}_s)} \left[ \mathcal{L}_{total} \right],
\end{equation}
where $\mathbb{E}$ denotes the expectation over the SNR distribution $p(\gamma)$ and the dataset distribution $p(\mathbf{H}_s)$, and $\mathcal{L}_{total}$ represents the composite loss function encompassing reconstruction error and regularization terms.

The VQ loss $\mathcal{L}_{vq}$ utilizes the stop-gradient operator $\text{sg}[\cdot]$ to constrain codebook updates and is formulated as
\begin{equation}
    \mathcal{L}_{vq} = \| \text{sg}[\mathbf{Z}] - \mathcal{C} \|_2^2 + \beta \| \mathbf{Z} - \text{sg}[\mathcal{C}] \|_2^2,
\end{equation}
where $\beta$ is a hyperparameter balancing the commitment loss.

\begin{figure*}[t]
    \centering
    \includegraphics[width=0.85\textwidth]{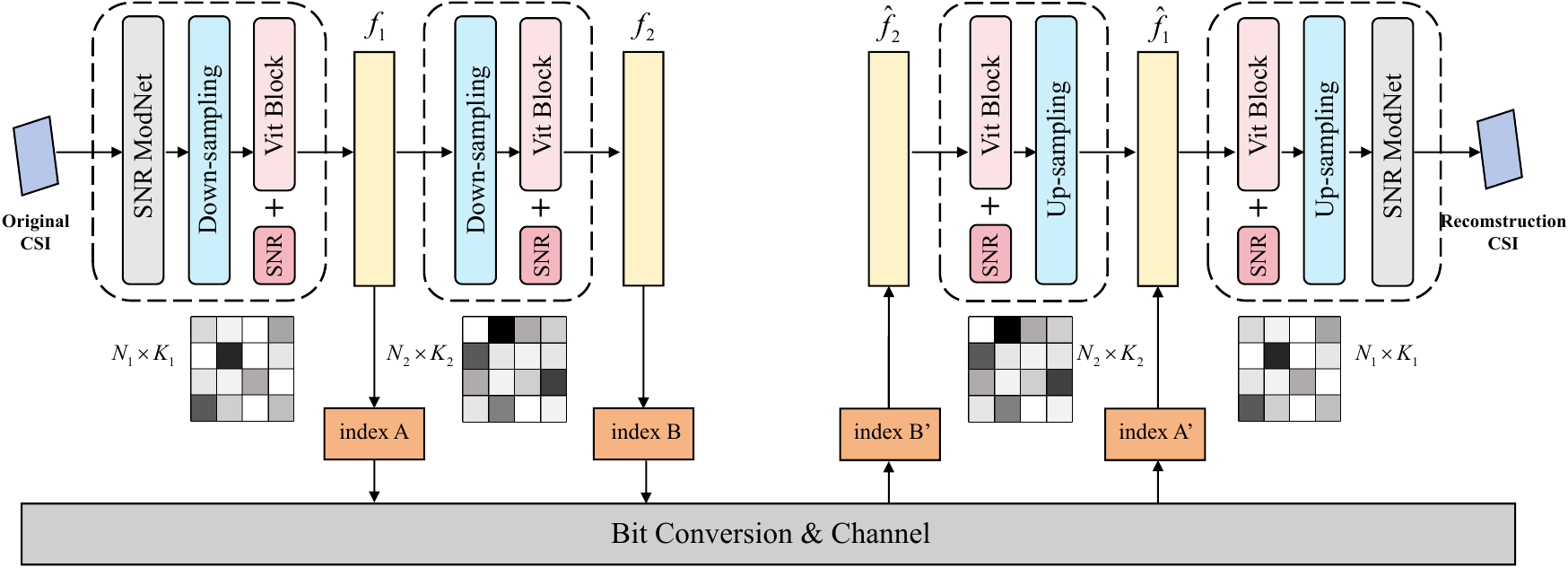} 
    \vspace{-0.5em}
    \caption{The Transformer-based architecture of the VQ-DJSCC-F.}
    \label{fig:network}
    \vspace{-1.5em}
\end{figure*}

Additionally, an entropy regularization term $\mathcal{L}_{ent}$ is introduced to maximize the uniformity of codeword usage. The total loss function can be expressed as:
\begin{equation}
    \mathcal{L}_{total} = \|\mathbf{H}_s - \hat{\mathbf{H}}_s\|_2^2 + \lambda_1 \mathcal{L}_{vq} + \lambda_2 \mathcal{L}_{ent},
\end{equation}
where $\lambda_1$ and $\lambda_2$ are weighting hyperparameters balancing the reconstruction quality, discrete representation stability, and codebook utilization. By jointly minimizing this composite objective, the network learns to extract robust and compact features under varying channel conditions.

\section{\scalebox{0.9}{PROPOSED DL-BASED CSI FEEDBACK METHOD}}
In this section, we detail the proposed VQ-DJSCC-F framework. The framework comprises four components: near-field data pre-processing, hierarchical VQ-VAE architecture, channel adaptation algorithms, and codebook training strategies.

\subsection{Channel Data Pre-processing with regard to near-field channel characteristics}

To achieve efficient CSI compression with limited feedback overhead, addressing the sparsity loss caused by near-field spherical wave propagation is a prerequisite. As discussed in Section II, the near-field effect leads to energy diffusion in the conventional angular domain. To counteract this, we leverage the joint sparsity inherent in the polar-delay domain to refocus the channel energy. Specifically, the original spatial-frequency domain channel matrix $\mathbf{H}$ is mapped into a low-dimensional sparse domain via a joint projection operation. The sparse matrix $\mathbf{H}_s$ can be expressed as
\begin{equation}
    \mathbf{H}_s = \mathbf{W}_{P} \mathbf{H} \mathbf{W}_{T}^H,
\end{equation}
where $\mathbf{W}_{P}$ denotes the angle-distance codebook matrix whose row vectors correspond to the basis vectors of a particular angle-distance binding, and $\mathbf{W}_{T} \in \mathbb{C}^{N \times N}$ represents the delay domain DFT matrix. Leveraging the physical property that multipath delays are concentrated within a finite time window, this transformation retains only the components containing valid energy. Through this processing, we directly extract the most energy-concentrated parts while effectively eliminating redundant noise, thereby enabling the subsequent VQ model to focus on learning critical features.

\subsection{Network Design}
To balance high-fidelity reconstruction with lightweight deployment, we design two architectures under the hierarchical VQ-DJSCC-F framework: a Transformer-based model prioritizing global feature modeling and a CNN-based model optimized for computational efficiency. Both employ multi-stage vector quantization but differ in feature extraction paradigms.

The Transformer-based architecture adopts a symmetrical hierarchical structure to capture long-range dependencies. The encoder uses strided convolutions to progressively downsample the input $\mathbf{H}_s$ into bottom-level ($16 \times 16$) and top-level ($8 \times 8$) feature maps. Feature extraction is performed by two cascaded Transformer Encoder Blocks. Specifically, each block consists of $L=4$ layers of multi-head self-attention ($N_{head}=4$) and feed-forward networks. To explicitly inform the network of channel conditions, an SNR Token is injected at the input of each block. Correspondingly, the decoding stage begins by refining the quantized top-level features using a 4-layer Transformer block. After up-sampling via transposed convolution, the resulting feature maps are integrated with the bottom-level quantized features through a $3 \times 3$ convolutional fusion layer. The fused features are processed by a second Transformer block and a final up-sampling layer to recover full-scale CSI.

The CNN-based architecture utilizes a lightweight pyramidal design. The encoder comprises three stages: Stage 1 extracts shallow features via $3 \times 3$ convolutions, while Stages 2 and 3 perform down-sampling using strided convolutions to generate bottom ($16 \times 16$) and top ($8 \times 8$) latent representations. To enhance robustness, each block integrates Batch Normalization, ReLU, and an SNR ModNet module. The decoder follows a cascaded reconstruction strategy, fusing up-sampled features with bottom-level representations for processing.

\subsection{SNR Adaptive Channel Mechanism}
To ensure robustness under time-varying channels, we design a hybrid channel adaptation strategy incorporating two mechanisms: SNR ModNet for feature calibration and SNR Token for attention guidance. These mechanisms operate at the spatial attention level.

SNR ModNet serves as a plug-and-play module, deployed after convolutional blocks in the CNN architecture and at the I/O interfaces of the Transformer architecture. It explicitly models channel-noise interdependencies using a dual-branch structure. The attention gating branch concatenates globally pooled features with an SNR embedding projected via an MLP, using a Sigmoid-activated layer to generate a channel-wise scaling factor. Parallel to this, a residual branch comprising $3 \times 3$ convolutional layers extracts local context. The final output is obtained by adding the residual features to the element-wise product of the input and the scaling factor, allowing the network to adaptively purify features based on instantaneous channel quality.

SNR Token is specifically tailored for the Transformer architecture to provide context-aware guidance. We project the scalar SNR into a high-dimensional embedding via an MLP consisting of Linear, LayerNorm, and ReLU layers, ensuring strict alignment with the latent feature dimension. During encoding and decoding, this token is concatenated as a prefix to the feature sequence. By participating in global self-attention, the SNR Token acts as a conditional prior influencing the Query-Key dot product calculation. This mechanism implicitly guides the Transformer to focus on dominant structural components, achieving robust adaptation.

\subsection{Codebook Updating and Entropy loss}
To address the "codebook collapse" issue common in VQ-VAE training and ensure high quantization precision, we employ an EMA update strategy and impose Entropy Regularization constraints.

Instead of relying on unstable gradient descent, we employ the EMA algorithm \cite{8} to smoothly update the codebook embeddings. Specifically, in each training step, we maintain two moving statistics: the accumulated cluster size $N_k^{(t)}$ and the accumulated embedding sum $\mathbf{M}_k^{(t)}$, utilizing a decay factor $\rho$. The update procedure is rigorously formulated as
\begin{align}
    N_k^{(t)} &= \rho N_k^{(t-1)} + (1-\rho) \sum_{j} \mathbb{I}(q_j^{(t)} = k), \\
    \mathbf{M}_k^{(t)} &= \rho \mathbf{M}_k^{(t-1)} + (1-\rho) \sum_{j} \mathbb{I}(q_j^{(t)} = k) \mathbf{z}_j^{(t)}, \\
    \mathbf{c}_k^{(t)} &= \frac{\mathbf{M}_k^{(t)}}{N_k^{(t)}}.
\end{align}
    where $\mathbb{I}(\cdot)$ denotes the indicator function, and $q_j^{(t)}$ represents the codebook index selected by the $j$-th feature vector $\mathbf{z}_j^{(t)}$ within the current batch. The first two lines update the count and vector sum of features assigned to the $k$-th codeword. The third line shows that the codeword $\mathbf{c}_k$ is updated by normalizing the accumulated embedding sum. This mechanism ensures the stability of codebook evolution.

Meanwhile, we introduce an entropy loss $\mathcal{L}_{ent}$ to maximize the representational capacity of the discrete feature space. This regularization term aims to maximize the entropy of codeword usage, thereby forcing the codebook probability distribution $p(\mathbf{c})$ towards a uniform distribution. The entropy loss is defined as
\begin{equation}
    \mathcal{L}_{ent} = \sum_{k=1}^K p_k \log p_k.
\end{equation}
where $p_k$ represents the usage frequency of the $k$-th codeword in the current batch. By minimizing this term to explicitly maximize the entropy, we effectively prevent the "collapse" phenomenon where certain codewords are overused while others remain idle, thereby significantly enhancing quantization efficiency.

\section{\scalebox{0.9}{SIMULATION RESULTS AND DISCUSSIONS}}
This section evaluates the performance of the proposed VQ-DJSCC-F system. The analysis is conducted within an AWGN channel environment. The near-field XL-MIMO dataset is generated based on the spherical wave propagation mechanism, configured with $N=256$ antennas, $M=1024$ subcarriers, and a central frequency of $f_c=100$ GHz. Users are randomly distributed within a range of 20 m to 60 m, strictly satisfying near-field conditions. Leveraging the joint sparsity, the pre-processing phase extracts a $32 \times 32$ effective feature matrix in the polar-delay domain. We partition the 150,000 samples into training, testing, and validation sets at a ratio of 10:3:2. The normalized mean square error (NMSE) is adopted as the core metric for performance evaluation.

The network training is implemented within the PyTorch framework using the Adam optimizer. The training process spans 100 epochs with a batch size of 100 and an initial learning rate of $2 \times 10^{-4}$, adjusted by a cosine annealing scheduler. To ensure adaptability, a mixed-SNR strategy uniformly samples SNR within $[0, 15]$ dB.

\vspace{-0.5em}
\begin{table}[ht]
    \centering
    \caption{Performance Comparison of Proposed Architectures under Different Feedback OVERHEADS}
    \label{table1}
    \renewcommand{\arraystretch}{1.1}
    \resizebox{\columnwidth}{!}{%
        \begin{tabular}{l|cccc}
            \hline
            \multicolumn{5}{c}{SNR$_{\text{train}} \in [0, 15]$ dB, SNR$_{\text{test}} = 0$ dB} \\
            \hline
            Feedback Overheads (Bits) & 1024 & 1536 & 1792 & 2048 \\
            \hline
            Proposed CNN (NMSE)  & -3.74 & -4.56 & -4.31 & -4.57 \\
            Proposed Transformer & -4.59 & -5.65 & -6.09 & -6.46 \\
            \hline
            \multicolumn{5}{c}{SNR$_{\text{train}} \in [0, 15]$ dB, SNR$_{\text{test}} = 15$ dB} \\
            \hline
            Feedback Overheads (Bits) & 1024 & 1536 & 1792 & 2048 \\
            \hline
            Proposed CNN (NMSE)  & -9.95 & -10.99 & -11.06 & -11.85 \\
            Proposed Transformer & -9.23 & -11.12 & -12.37 & -13.17 \\
            \hline
        \end{tabular}%
    }
\end{table}

Table \ref{table1} presents the baseline reconstruction performance of the proposed CNN-based and Transformer-based architectures across varying feedback overheads. The results demonstrate a monotonic improvement in reconstruction accuracy for both models as the feedback overhead increases, under both low and high SNR conditions. This validates the effectiveness of the framework in diverse feedback scenarios.

Notably, the Transformer-based architecture exhibits superior reconstruction fidelity and generalization capabilities. By leveraging its inherent strength in modeling global dependencies, the Transformer consistently outperforms CNN. This advantage is particularly pronounced in low-SNR regimes. Moreover, the performance gap widens as the feedback overhead increases, underscoring the Transformer's superior efficacy in utilizing additional bits to recover high-frequency details. In contrast, while the CNN-based architecture presents a slightly lower performance ceiling, it maintains competitive accuracy with significantly reduced computational complexity. This trade-off enables the proposed VQ-DJSCC-F framework to accommodate diverse deployment constraints flexibly.

The Transformer-based model is trained over a uniform SNR range of $[0, 15]$ dB. For benchmarking purposes, we employ three VQ-DJSCC-F models lacking the adaptive modules, each trained specifically at one fixed SNR$_{\text{train}} $ level ($0\text{dB}$, $6\text{dB}$, or $12\text{dB}$). Furthermore, to provide a comprehensive evaluation, we introduce a traditional separate source-channel coding (SSCC) benchmark: the 5G NR Type II codebook concatenated with Polar codes. Digital communication is implemented using QPSK modulation alongside OFDM.

\begin{figure}[h]
    \centering
    \includegraphics[width=0.75\linewidth, keepaspectratio]{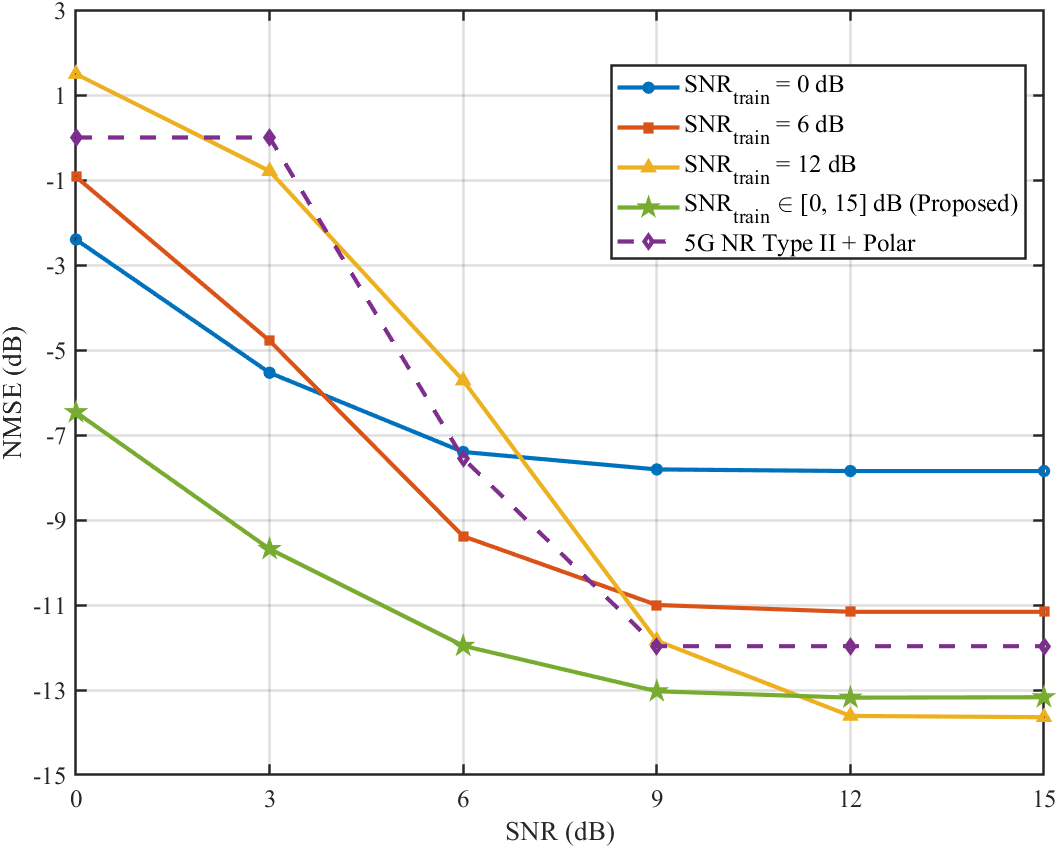}
    \caption{ Channel adaptive performance of VQ-DJSCC-F.}
    \label{fig1}
\end{figure}

Fig. \ref{fig1} compares the NMSE performance over AWGN channels. Observations indicate: 1) Unlike the 5G method, which suffers from the ``cliff effect'' and performance collapse at low SNRs due to Polar codes' threshold characteristics, the proposed VQ-DJSCC-F demonstrates smooth performance degradation. This characteristic ensures superior robustness in the low-SNR regime. 2) The standalone models lacking adaptive modules suffer from severe channel mismatch. In contrast, the proposed model maintains near-optimal performance across the entire dynamic range by leveraging SNR adaptation strategies. 3) A higher $\text{SNR}_{\text{test}}$ monotonically improves the performance of VQ-DJSCC-F. Crucially, regardless of the deviation between $\text{SNR}_{\text{train}}$ and $\text{SNR}_{\text{test}}$, the proposed method maintains a significant advantage over the baseline without the SNR ModNet.

\begin{figure}[h]
    \centering
    \vspace{-0.2cm}
    \includegraphics[width=0.75\linewidth, keepaspectratio]{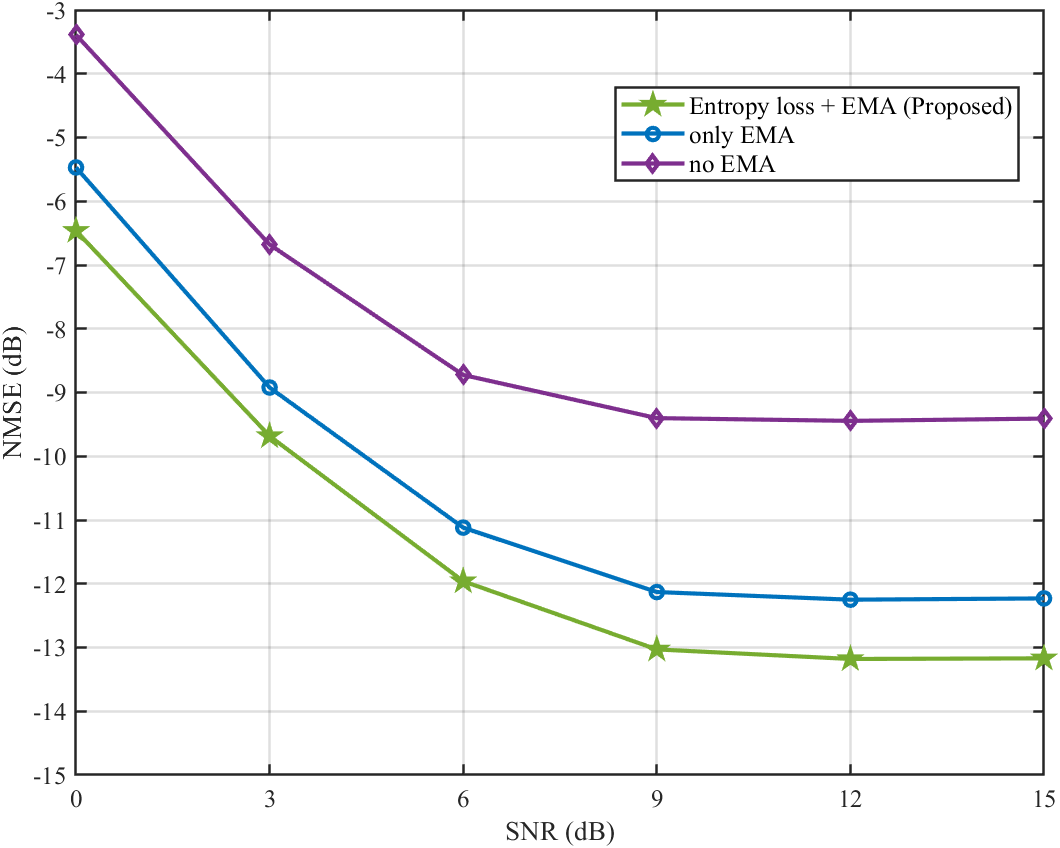}
    \caption{ Performance of VQ-DJSCC-F under different methods.}
    \label{fig2}
    \vspace{-0.2cm}
\end{figure}

Fig. \ref{fig2} presents the ablation study on codebook optimization strategies. We evaluate the NMSE performance under three configurations: the proposed method, a variant utilizing only EMA without the Entropy loss, and a baseline using standard gradient updates that excludes both Entropy loss and EMA. In contrast, introducing EMA significantly stabilizes updates and boosts performance. Furthermore, the proposed method outperforms the EMA-only variant across the entire SNR range. This confirms that combining EMA with entropy maximization effectively mitigates codebook collapse and enhances codebook utilization, ensuring optimal performance.

\section{Conclusion}
This paper proposed a digital CSI feedback framework tailored for 6G XL-MIMO systems, named \mbox{VQ-DJSCC-F}. By exploiting near-field sparsity in the polar-delay domain, we achieved effective dimension reduction and energy-concentrated feature extraction. We designed a hierarchical architecture with flexible Transformer and CNN backbones to encode features into a discrete latent space. To ensure high-precision quantization and transmission robustness, we introduced a joint entropy-EMA codebook optimization strategy alongside an attention-driven channel adaptation mechanism. Simulation results demonstrate that the proposed scheme significantly outperforms benchmarks, achieving superior reconstruction accuracy and stability under varying channel.

\bibliographystyle{ieeetr}
\bibliography{Mybib}

\end{document}